# The Role of Terahertz and Far-IR Spectroscopy in Understanding the Formation and Evolution of Interstellar Prebiotic Molecules

**Duncan V. Mifsud[1,2*], Perry A. Hailey[1], Alejandra Traspas Muiña[3], Olivier Auriacombe[4], Sergio Ioppolo[3], Nigel J. Mason[1]**

[1]Centre for Astrophysics and Planetary Science, School of Physical Sciences, University of Kent, Canterbury CT2 7NH, United Kingdom

[2]Institute for Nuclear Research (Atomki), Debrecen H-4026, PO Box 51, Hungary

[3]School of Electronic Engineering and Computer Science, Queen Mary University of London, London E1 4NS, United Kingdom

[4]Microwave Electronics Laboratory, Department of Microtechnology and Nanoscience, Chalmers University of Technology, Göteborg 41296, Sweden

**\* Correspondence:**
Duncan V. Mifsud
duncanvmifsud@gmail.com

**Abstract**

Stellar systems are often formed through the collapse of dense molecular clouds which, in turn, return copious amounts of atomic and molecular material to the interstellar medium. An in-depth understanding of chemical evolution during this cyclic interaction between the stars and the interstellar medium is at the heart of astrochemistry. Systematic chemical composition changes as interstellar clouds evolve from the diffuse stage to dense, quiescent molecular clouds to star-forming regions and proto-planetary disks further enrich the molecular diversity leading to the evolution of ever more complex molecules. In particular, the icy mantles formed on interstellar dust grains and their irradiation are thought to be the origin of many of the observed molecules, including those that are deemed to be 'prebiotic'; that is those molecules necessary for the origin of life. This review will discuss both observational (e.g., ALMA, SOFIA, Herschel) and laboratory investigations using millimeter, submillimeter, and terahertz and far-IR (THz/F-IR) spectroscopies and the role that they play in contributing to our understanding of the formation of prebiotic molecules. Mid-IR spectroscopy has typically been the primary tool used in laboratory studies. However, THz/F-IR spectroscopy offers an additional and complementary approach in that it provides the ability to investigate intermolecular interactions compared to the intramolecular modes available in the mid-IR. THz/F-IR spectroscopy is still somewhat under-utilized, but with the additional capability it brings, its popularity is likely to significantly increase in the near future. This review will discuss the strengths and limitations of such methods, and will also provide some suggestions on future research areas that should be pursued in the coming decade exploiting both space-borne and laboratory facilities.

# 1    Interstellar Chemistry: From Principles to Practice

The idea of active chemistry occurring within interstellar space is a relatively new one. Historically, interstellar temperatures and pressures were considered to be far too low to permit atoms or ions to come into contact with each other and overcome a reaction activation energy barrier. However, since the first evidence of interstellar molecular species arose in the late 1930s, the scientific community has come to recognize that the interstellar medium is home to a plethora of structurally diverse molecules, with different environments displaying their own characteristic chemistry. A more complete description of the history of astrochemistry may be found in the work of Feldman (2001).

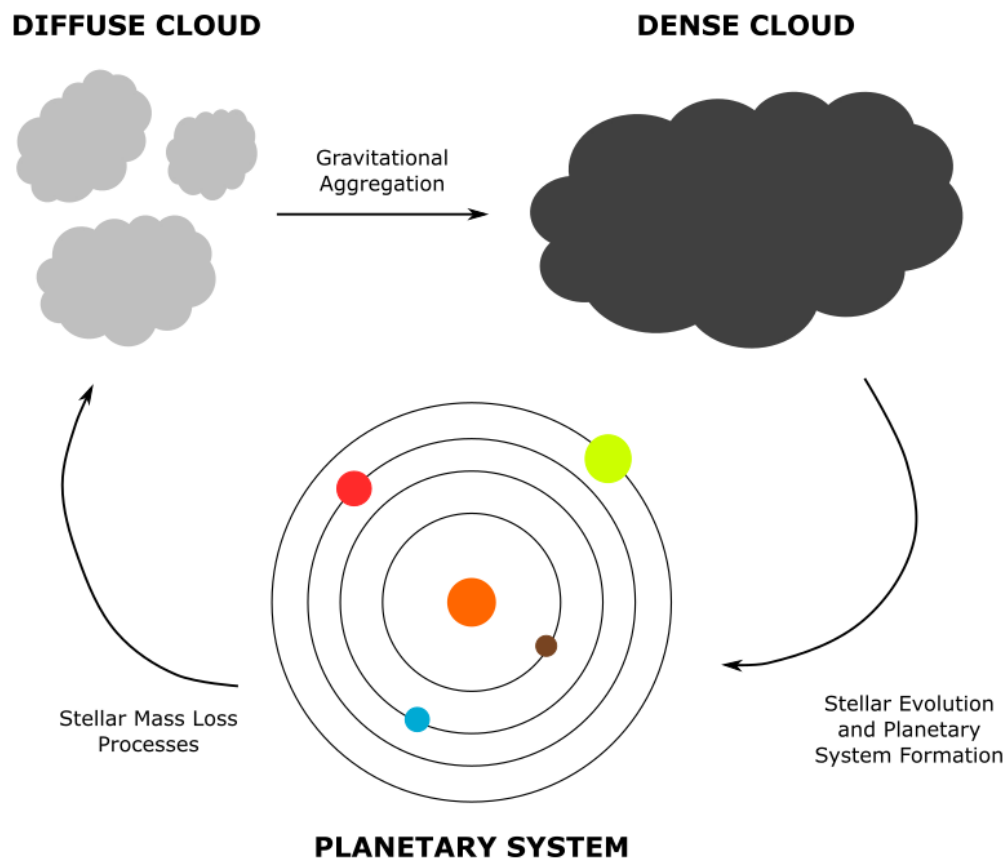

**Figure 1** Matter in the interstellar medium is cycled through various phases: the aggregation of matter results in optically transparent diffuse clouds evolving into dense, dark molecular clouds. These clouds are considered to be 'stellar nurseries' in which localized gravitational collapse results in the formation of a proto-star, with further accretion of material occurring via magneto-hydrodynamic flows onto the growing star. Within a few million years, the stellar object evolves into a pre-main sequence star surrounded by a proto-planetary disk. Collisions within the disk result in the accretion of dust grains which slowly grow to form planetesimals, thus taking the first steps towards the formation of a planetary system. Once the star's nuclear fuel has been exhausted, they may follow one of two routes depending on the stellar mass: low-mass stars undergo several mass-loss processes in which molecular material is dispersed back into the interstellar medium leaving behind a remnant white dwarf, whereas high-mass stars instead explode as supernovae with their cores eventually becoming either neutron stars or black holes.

Matter in the interstellar medium is composed of around 99% atomic or molecular hydrogen (H or $H_2$) and 1% carbonaceous or silicate dust grains. This matter is cycled through a number of phases each displaying a characteristic chemistry (Fig. 1). The smallest over-densities result inf the formation of diffuse clouds, which possess particle densities of $10^2$-$10^3$ cm$^{-3}$ and kinetic temperatures of a few hundred K (Smith 1992). Diffuse clouds are relatively transparent to visible light and are continuously exposed to the interstellar radiation field which results in the dissociation and ionization of molecular material. As such, these clouds are dominated by the presence of atoms and ions, particularly H and $H^+$, although a few molecular species do survive (Snow and Bierbaum 2008). Indeed, the first molecules to be detected in interstellar space were detected in diffuse interstellar media via optical spectroscopy (Smith 1992, Feldman 2001, Larsson et al. 2012). With the notable exception of the formation of $H_2$ (Wakelam et al. 2017), most of the chemistry within diffuse interstellar clouds is mediated by gas-phase reactions (Geppert and Larsson 2013).

Accumulation of gaseous material within diffuse clouds may occur as a result of random perturbations, stellar winds, or more violent phenomena such as supernovae-induced shock waves. As diffuse clouds begin to accumulate matter, the proportion of molecular material (mainly $H_2$) increases. This is accompanied by a diminishing of the proportion of ionized species present: for example, the dominant form of carbon changes to the atomic form from $C^+$ and the number of free electrons also decreases (Snow and Bierbaum 2008). Thus, the accumulation of matter results in an evolution of so-called 'atomic' diffuse clouds to 'molecular' diffuse clouds.

Further accumulation of material as a result of gravitational attractions results in the formation of a dense interstellar cloud (also known as a dark interstellar cloud). These clouds are, as their name suggests, characterized by higher particle number densities of up to $10^6$ cm$^{-3}$ and are the primary location of stellar formation. The increase in density results in a concomitant increase in cloud opacity. As such, the far- and vacuum-UV components of the interstellar radiation field are attenuated at the edges of the cloud, resulting in an interior with a temperature of 7-20 K (Galli et al. 2002). The higher particle densities of these dark clouds enhance the efficiency of reactions in the gas phase, particularly those between ions and molecules, and contribute to the formation of several new species including complex organic molecules (Bergin and Tafalla 2007, Öberg 2016, Arumainayagam et al. 2019).

At the low temperatures encountered within the interiors of dense interstellar clouds, many species are able to freeze-out onto the surfaces of the carbonaceous and silicate dust grains, thus forming an icy mantle. This icy mantle is typically composed of two zones (Fig. 2): a lower polar layer which results from the deposition and synthesis (via hydrogenation reactions) of polar molecules such as $H_2O$, $NH_3$, and $CH_4$; and an upper apolar layer characterized by the presence of CO and related molecules such as $CO_2$ and $CH_3OH$ (Larsson et al. 2012, Öberg 2016).

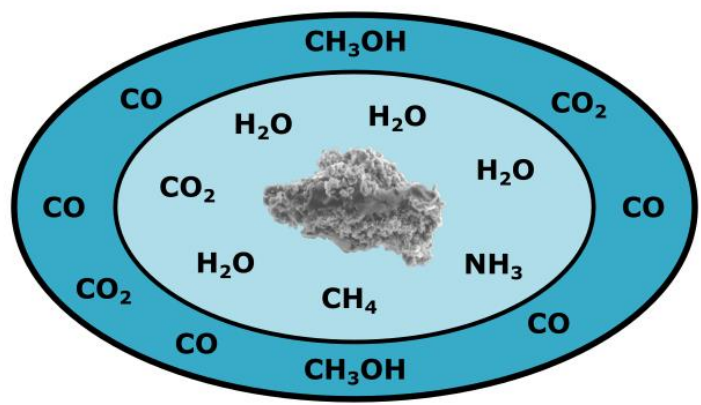

**Figure 2** Median structure of icy grain mantles within dense interstellar clouds. Shown in the image, a diagram (not to scale) of a typical icy mantle divided into two zones: a lower polar layer rich in $H_2O$ molecules and hydrogenated heteroatoms, and an upper apolar layer largely consisting of CO and related molecules. More detailed information may be found in the work of Öberg (2016).

These molecular icy grain mantles play host to a rich chemistry mediated by several phenomena and which may result in the formation of complex molecules. For instance, galactic cosmic rays are not attenuated by the boundaries of the dense cloud, and so are able to penetrate through to the cloud core and induce chemistry within the icy layers. Interactions between these cosmic rays and gas-phase $H_2$ results in the emission of Lyman-$\alpha$ photons which may induce photochemistry. Photochemistry in icy grain mantles may also be induced by lower wavelength photons which enter the cloud via less dense regions. Thermal chemistry may also occur in ices proximal to a heat source, such as a nascent star.

In this section, we review the chemistry which occurs within diffuse and dense interstellar clouds with a particular focus on reactions involving ionic species, complex organic molecules. Although mention of detections via terahertz and far-IR (THz/F-IR) spectroscopic techniques will be made, further details on such techniques will be given in the following section. Reviews on the chemistry occurring during the various other phases of the astrochemical cycle (Fig. 1) are available elsewhere (van Dishoeck and Blake 1998, Williams and Hartquist 1999, Larsson et al. 2012, Geppert and Larsson 2013).

## 1.1 Diffuse Interstellar Clouds: Ion Chemistry and THz/F-IR Spectroscopy

As mentioned previously, matter within the most diffuse interstellar clouds is primarily composed of atomic and ionic material, with a significantly smaller number of molecules also being present due to the high rates of photon and cosmic ray induced dissociation and ionization processes. Consequentially, the chemistry within these clouds is dominated by gas-phase reactions including those involving or producing ions. At the edges of the diffuse cloud, where stellar UV irradiation is more intense, photo-ionization of carbon atoms produces $C^+$, while within the diffuse cloud interior cosmic ray driven chemistry is more dominant and allows for the ionization of H and $H_2$ (Smith 1992). The formation of $H_2^+$ in particular is important, as this species is the direct precursor to $H_3^+$:

$$H_2 + H_2^+ \rightarrow H + H_3^+$$

For a long time, the formation and presence of $H_3^+$ within diffuse clouds were highly controversial due to its suspected high rate of dissociative recombination mediated by electrons emitted from the photo-ionization of carbon atoms (Larson et al. 2000, Larsson et al. 2008). However, its direct observation in diffuse clouds (McCall et al. 1998) combined with the discovery of a new method of recombination (Kokoouline and Greene 2003) confirmed its presence and role within interstellar chemistry: $H_3^+$ efficiently donates a proton to other atoms and molecules due to the low proton affinity of $H_2$. The resultant protonated species is reactive, and is thus capable of triggering various chains of ion-molecule reactions. For instance, the reaction with CO produces $HCO^+$, while the reaction with atomic oxygen produces $OH^+$:

$$H_3^+ + CO \rightarrow H_2 + HCO^+$$

$$H_3^+ + O \rightarrow H_2 + OH^+$$

$OH^+$ may also be formed as a result of a two-step processes starting with the charge-transfer reaction between atomic oxygen and $H^+$ to yield $O^+$, which subsequently reacts with $H_2$ to furnish $OH^+$ (Yamamoto 2017). The formation of $OH^+$ is the first step towards the formation of interstellar gas-phase $H_2O$. The reaction of $OH^+$ with one moiety of $H_2$ yields $H_2O^+$, with the reaction with a second moiety producing $H_3O^+$ (Larsson et al. 2012). The $H_3O^+$ molecular ion is isoelectronic with $NH_3$ and so is a stable species which resists further reactions with $H_2$ (Geppert and Larsson 2013). Instead, it undergoes dissociative recombination with electrons to yield either OH (~60 of outcomes) or $H_2O$ (~25% of outcomes).

The detections of $OH^+$, $H_2O^+$, and $H_3O^+$ within diffuse interstellar environments have been made using ground- and space-based telescopes working in the THz/F-IR range such as APEX and the HIFI instrument aboard the Herschel Space Observatory (for more detail, refer to section 3), thus confirming their contribution to chemistry in such settings (Wyrowski et al. 2010, Gerin et al. 2010a, Gupta et al. 2010, Wiesemeyer et al. 2016). A reaction scheme analogous to the formation of $H_2O$ from $OH^+$ has also been invoked to explain the formation of HCl within diffuse clouds starting from $Cl^+$. Such a reaction scheme has been validated by the detection of the relevant intermediate species $HCl^+$ and $H_2Cl^+$ by the HIFI instrument aboard the Herschel Space Observatory (Lis et al. 2010, De Luca et al. 2012).

Reactions leading to the formation of carbon-bearing molecules within diffuse interstellar media are more challenging to explain compared to the oxygen and chlorine chemistry outlined above. Indeed, the formation of $CH^+$, one of the earliest known interstellar molecules (Douglas and Herzberg 1941), is still not completely understood and several simulations have underestimated its observed abundance within diffuse clouds (e.g., van Dishoeck and Black 1986, Pan et al. 2004, Godard et al. 2009). It is apparent that the formation of $CH^+$ cannot rely on the reaction between $H_2$ and $C^+$ due to the considerable endothermicity of this reaction (Myers et al. 2015, Valdivia et al. 2017, Moseley et al. 2021, Plašil et al. 2021). In light of this, alternative mechanisms have been suggested to account for its presence within the diffuse interstellar medium. For instance, it has been suggested that the radiative association of $C^+$ and $H_2$ to yield $CH_2^+$ with the concomitant emission of a photon is the first step towards interstellar $CH^+$, as the photo-dissociation of the former may lead to the latter (Yamamoto 2017). However, the rate at which radiative association occurs is fairly low (Graff et al. 1983, Barinovs and van Hemert 2006) and indeed the primary reaction pathway of $CH_2^+$ within diffuse interstellar media is the reaction with $H_2$ which furnishes $CH_3^+$. As such, the potential contribution of this reaction network in accounting for interstellar $CH^+$ abundances is probably low.

Another proposed formation route for $CH^+$ is the direct reaction between $C^+$ and $H_2$ if the endothermicity of the reaction is overcome by some mechanism which causes a localized warming effect which raises the temperature to a few thousand K. Initially, it was thought that this localized warming could be achieved in regions of the diffuse cloud processed by magneto-hydrodynamic shock waves (Elitzur and Watson 1980, Draine and Katz 1986). However, such processes are now thought to be an unlikely source of $CH^+$ as the radial velocity for this molecule and for CH have been observed to be similar. Among other ideas, gas heating caused by turbulences within the cloud has also been found to be unsatisfactory in accounting for the observed abundances of $CH^+$ (Godard and Cernicharo 2013), as have reactions between $C^{2+}$ and $H_2$ in X-ray dominated regions of the diffuse cloud (Plašil et al. 2021). Alfvén waves have also been suggested as a potential source of $CH^+$ (Federman et al. 1996), however, to the best of the authors' knowledge, this has been neither corroborated nor firmly rejected. As such, the elusive source of $CH^+$ within diffuse interstellar media remains an unanswered question. There thus exists a strong motivation to address this problem, as the formation of $CH^+$ may be an important step in the synthesis of significantly more complex carbon-based molecules such as fullerenes and polycyclic aromatic hydrocarbons (PAHs), which are suspected of being the primary carriers of observed diffuse interstellar bands and which may bear relevance to the synthesis of prebiotic molecules (Taylor and Duley 1997).

With regards to molecular detections in diffuse interstellar clouds, a significant number have been made using methods and techniques which exploit the radio region of the electromagnetic spectrum, including THz/F-IR spectroscopy (Snow and McCall 2006, Yamamoto 2017). Indeed, such techniques are the preferred method for detecting polyatomic molecules, as these molecules are difficult to identify using optical spectroscopy (as was done historically) due to the pre-dissociative nature of their excited states. The HIFI instrument aboard the Herschel Space Observatory in particular has proven to be invaluable in the detection and mapping of molecules in diffuse interstellar media, having provided much information on the chemistry of hydrogen-, carbon-, oxygen-, nitrogen-, sulfur-, fluorine-, and chlorine-bearing molecules within diffuse clouds (Sonnentrucker et al. 2010, Langer et al. 2010, Persson et al. 2010, Neufeld et al. 2010a, 2010b, 2012a, Gerin et al. 2010a, 2010b, 2012, Monje et al. 2011, 2013, De Luca et al. 2012, Esplugues et al. 2014, Schilke et al. 2014, Indriolo et al. 2015, Jacob et al. 2020).

Other instruments and telescopes have also aided greatly in the elucidation of the chemical complexity of the diffuse interstellar medium. For example, APEX, ALMA, and the GREAT instrument aboard SOFIA have all been used in the detection of sulfur-bearing molecules, including isotopologues (Menten et al. 2011, Neufeld et al. 2012b, 2015, Muller et al. 2017). Perhaps the most exciting of all has been the series of recent ALMA observations of complex molecules. The work of Thiel et al. (2017), for instance, investigated the presence of prebiotic complex organic molecules within diffuse media and detected several species, including $CH_3OH$, $CH_3CN$, $CH_3CHO$, $HC_3N$, and $CHONH_2$. Such results are of immense consequence, as they imply that certain molecular precursors to life either form early on in the astrochemical cycle (Fig. 1), or else survive a process of expansion from denser interstellar media.

## 1.2 Dense (Dark) Interstellar Clouds: Ion Chemistry and THz/F-IR Spectroscopy

Condensation of diffuse clouds results in the formation of dense (dark) clouds. During this process, the dominant material is converted from atomic and ionic species to molecular ones. For instance, $H_2$ becomes the dominant form of hydrogen, while the dominant form of carbon changes sequentially from $C^+$, to C, and finally to CO (Snow and Bierbaum 2008, Larsson et al. 2012). As such, these structures have also been appropriately dubbed molecular clouds. In their review, Arumainayagam et al. (2019) highlighted the chemical productivity of dense clouds and identified three milieux for astrochemical reactions within them. The first of these is gas-phase chemistry, which includes radical-radical, radical-neutral, and ion-neutral reactions. The second is chemistry occurring on bare carbonaceous or silicate dust grain surfaces, while the third includes all chemical reactions occurring via energetic and non-energetic processing of icy grain mantles. In this section of the review, we will largely limit ourselves to a brief discussion of the ion-molecule reactions which occur in the gas phase, and the use of THz/F-IR spectroscopy in this aspect of the science.

Gas-phase ion-neutral reactions within dense interstellar clouds are favorable, barrierless chemical processes. The source of the ionic species within the cloud is somewhat dependent upon the location of the reaction within the cloud structure. Within the densest regions (surrounding the cloud core), ionization of molecules is caused almost exclusively by penetrating galactic cosmic rays. However, within the less dense regions of the clumpy cloud structure, far- and vacuum-UV components of the interstellar radiation field are the primary cause of molecular ionization via photon-induced electron loss processes. Furthermore, in the presence of OB stars, irradiation of molecular material by X-rays may also contribute to the formation of ionic species.

Cosmic ray induced ionization of $H_2$ leads indirectly to the formation of $H_3^+$ as was described for diffuse interstellar clouds. The low proton affinity of $H_2$ allows $H_3^+$ to be an efficient proton donor to other species (Larsson et al. 2012). For example, the reaction with $N_2$ yields $N_2H^+$, which is frequently used as a tracer for $N_2$ within interstellar space when observing with radio telescopes, since $N_2$ does not possess a permanent dipole moment. Ionized hydrogen, primarily in the form of $H^+$, readily undergoes reactions with other atoms such as nitrogen atoms. The resultant product $NH^+$ may undergo subsequent hydrogenation reactions with $H_2$ to sequentially form $NH_2^+$, $NH_3^+$, and $NH_4^+$ (Geppert and Larsson 2013). Finally, dissociative recombination with an incident electron efficiently affords gas-phase $NH_3$ (Öjekull et al. 2004).

Interestingly, the analogous reaction pathway to the formation of gas-phase $CH_4$ is not thought to be significant within dense clouds. This is because once $CH_5^+$ is formed, the dominant dissociative recombination mechanism involves the simultaneous loss of two hydrogen atoms, creating $CH_3$ rather than $CH_4$ (Semaniak et al. 1998, Viti et al. 2000). The formation of $CH_4$ and several other carbon-based molecules within the interstellar medium is instead likely to rely on other mechanisms, such as hydrogenation reactions at the surfaces of dust grains.

A large number of carbon-based complex organic molecules are known to exist within dense clouds. Until recently, it was thought that these molecules formed within the solid phase in icy grain mantles and were released into the gas phase upon thermal or photo-desorption. However, a growing body of literature has documented the presence of such gas-phase complex organic molecules within starless and pre-stellar cores, where such desorption processes do not take place. Scibelli and Shirley (2020), for instance, have detected $CH_3OH$ and $CH_3CHO$ in several such environments and have suggested that the radicals and molecules responsible for the formation of these species may be formed in the ice phase. The release of energy upon their formation allows them to sublimate into the gas phase (a process termed reactive desorption) whereupon they react to form these product molecules.

Another region of dense clouds which may host a rich gas-phase chemistry leading to the formation of complex organic molecules is the edge of the cloud: the photo-dissociation region (PDR). In this region, the cloud interacts with the intense UV components of the interstellar radiation field resulting in the desorption of solid-phase molecules and an increase in the occurrence of gas-phase processes. Using the IRAM 30 m telescope, Cuadrado et al. (2017) were able to perform a millimeter line survey towards the edge of the Orion Bar PDR and successfully detected a plethora of complex molecules including carboxylic acids, aldehydes, alkenes, and sulfur-containing molecules. Soma et al. (2018) reported similar results. The production and survival of complex molecules in such an environment is significant from the perspective of astrobiology and, although their exact formation mechanisms are still not completely understood, it is likely that a non-negligible contribution comes from gas-phase reactions.

The presence and distribution of complex, potentially prebiotic molecules within the interstellar medium have been extensively probed using various ground- and space-based observatories. Perhaps the most productive of these studies, however, have been those which have made use of the ALMA for molecular detections (Belloche et al. 2016, 2019, Garrod et al. 2017, Jørgensen et al. 2018, Calcutt et al. 2019, Willis et al. 2020). The impact of ALMA observations on the detection of interstellar molecules has been massive: first detections of prebiotic molecules which were previously controversial or unconfirmed, such as urea ($NH_2CONH_2$), have been achieved (Belloche et al. 2019), and more information has been gleaned on interstellar chemical processes such as deuteration (Belloche et al. 2016).

To conclude this section, we provide an overview of two especially fruitful ongoing observational projects which include THz/F-IR frequencies in their operational ranges: the GOTHAM and ARKHAM projects. These projects are run using the Robert C. Boyd 100 m Green Bank Telescope in the United States and have provided significant data demonstrating the presence of complex organic molecules within interstellar space, with a particular focus on aromatic species. Although operating in a similar fashion, the projects themselves have somewhat different objectives.

The GOTHAM project is primarily concerned with establishing a chemical inventory of the Taurus Molecular Cloud, one of the nearest dense clouds, by performing high-sensitivity wide-band spectral line surveys. Such work is motivated by the fact that benzonitrile ($c$-$C_6H_5CN$) was recently discovered in this pre-stellar source (McGuire et al. 2018a), contrasting with a working hypothesis which postulates that large interstellar aromatic molecules are produced in the circumstellar regions of post-AGB stars and subsequently break down into smaller sub-units such as benzene ($c$-$C_6H_6$) under energetic processing (Tielens 2008): so-called 'top-down' chemistry. However, these results suggest that reactions between smaller precursor molecules may also contribute to the formation of larger aromatics within pre-stellar regions through a 'bottom-up' chemistry approach.

GOTHAM observations of the Taurus Molecular Cloud, combined with laboratory-generated spectra and computational models, have successfully identified many molecules, some of which had previously not yet been described in the interstellar medium. The unsaturated molecules detected within this dense cloud may play a pivotal role in the formation of larger aromatic molecules, as well as other species relevant to the origins of life. Early work from GOTHAM identified propargyl cyanide ($HCCCH_2CN$) and the cyanopolyyne $HC_{11}CN$ in the Taurus Molecular Cloud (McGuire et al. 2020, Loomis et al. 2021). Cyclic molecules have also been discovered, including 1-cyanocyclopentadiene and 2-cyanocyclopentadiene (McCarthy et al. 2021, Lee et al. 2021a). The project has also definitively identified the first individual PAH species in interstellar space; 1-cyanonaphthalene, 2-cyanonaphthalene, and indene (McGuire et al. 2021, Burkhardt et al. 2021a). Other cyano and isocyano species have also been detected (Xue et al. 2020, Lee et al. 2021b).

The ARKHAM project is focused on investigating whether the chemistry observed in the Taurus Molecular Cloud is also found in other pre-stellar and proto-stellar sources so as to assess whether or not the resultant molecules survive the birth and evolution of a protostar. Results from the project have shown that benzonitrile has been detected in four additional sources: Serpens 1A, Serpens 1B, Serpens 2, and MC27/L1521F (Burkhardt et al. 2021b). Such detections illustrate that aromatic chemistry is favorable and extensive throughout the very earliest stages of star formation. Interestingly, the observed abundances of benzonitrile exceeded those predicted by models, despite these models being well-suited to other large carbon-based molecules such as cyanopolyynes. It is therefore probable that the formation mechanisms for aromatic species in pre-stellar sources differ from those of linear molecules and are not yet well understood.

### 1.3 The Link to Prebiotic Chemistry

Joint observational and laboratory investigations are key to deciphering the exact mechanisms of formation of the various complex organic molecules observed within interstellar and circumstellar media. To date, these studies have largely relied on a 'bottom-up' chemical perspective in which large complex molecules are produced from simpler starting molecules undergoing some form of processing (such as photolysis or radiolysis). However, as has been demonstrated by the GOTHAM and ARKHAM projects, the formation of certain molecules (such as aromatic species) is likely to be dependent on some combination of 'bottom-up' and 'top-down' chemistry (McGuire et al. 2020): that is to say that they are formed both from the joining together of smaller molecules, ions, and radicals, as well as the break-up of larger structures (Fig. 3).

PAHs are perhaps the best class of interstellar molecules to exemplify the contributions of 'top-down' and 'bottom-down' astrochemistry due to the fact that they are relatively ubiquitous in interstellar media, and also because they are able to undergo many chemical transformations yielding a plethora of molecular structures with various sizes, complexities, and arrangements of the aromatic rings, as well as different chemical characteristics such as the inclusion of heteroatoms and the addition of functional groups (such as amino or carboxylic moieties). Indeed, such chemical transformations are thought to be key to the formation of very small grains and fullerenes which result from the respective agglomeration and destruction of PAHs, as well as serving as a carbon feedstock for the production of molecules relevant to biology (Tielens 2008, Groen et al. 2012).

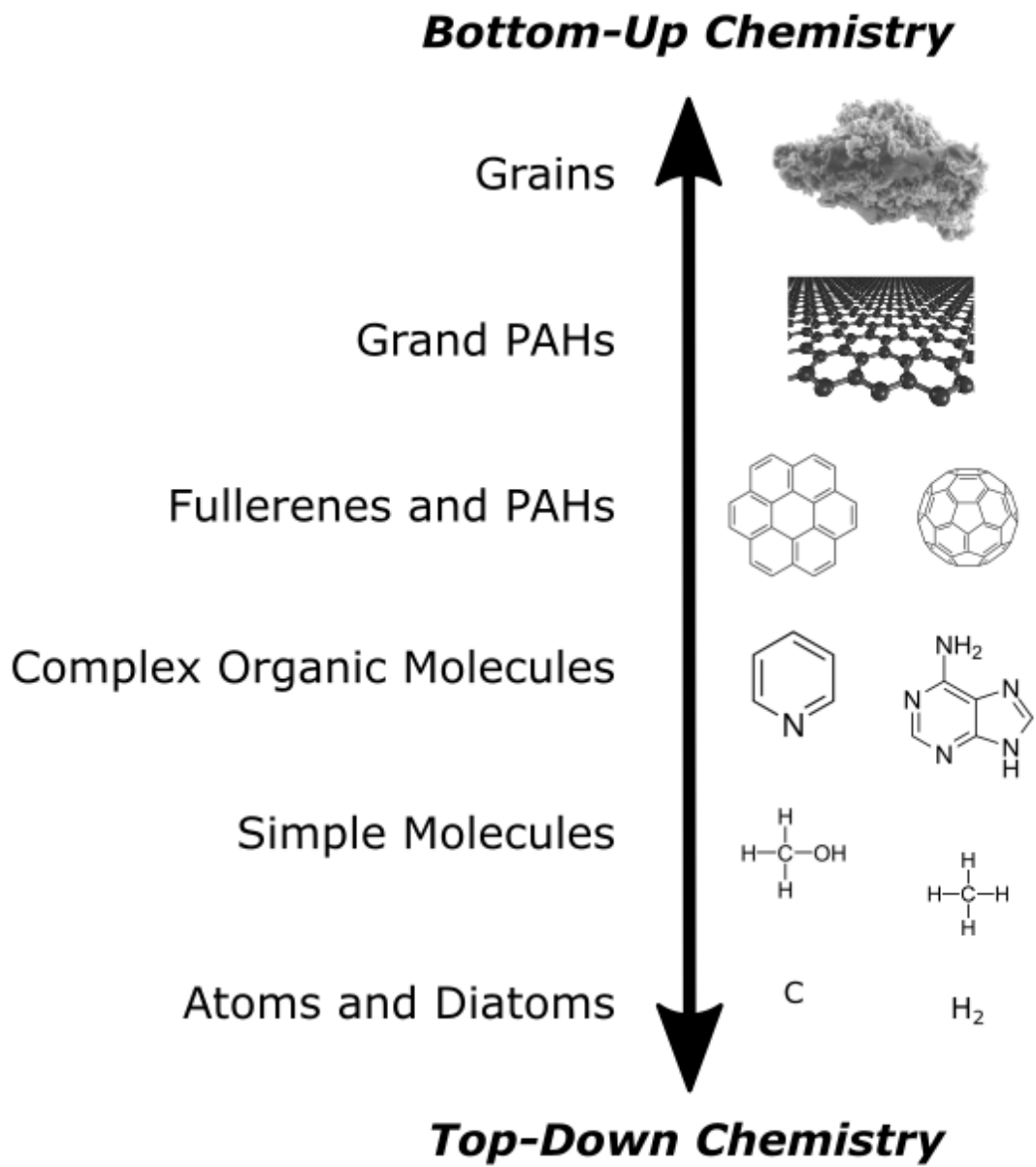

**Figure 3** Schematic diagram exhibiting models of 'top-down' and 'bottom-up' chemistry using carbon molecules as an example. In the 'bottom-up' perspective, the association of atoms and simple molecules allows for the formation of larger molecules such as $CH_4$ and $CH_3OH$. Further additions to the molecular structure may result in the formation of a complex organic molecule, such as an aromatic molecule or a PAH. Two-dimensional extension of the PAH structure produces graphite, which itself may be considered the basis of (sub-)micron sized interstellar carbonaceous dust grains. The 'top-up' chemical perspective is essentially the reverse of this, in which larger molecular structures are broken up as a result of energetic processing, thus yielding simpler molecules.

For instance, when mixed with $H_2O$ and subjected to energetic processing (a scenario which is plausible within icy grain mantles in the dense interstellar medium), PAHs have been shown to incorporate oxygen into their molecular structure and display new functional groups such as alcohols, ethers, and ketones (Cook et al. 2015). The addition of a ketone moiety to the PAH is biologically significant, as this provides the base structure for the vitamin K chemical family. Incorporation of heteroatoms (such as oxygen and nitrogen) into the ring structure of the PAH as a result of the energetic processing of mixed icy grain mantles is also possible (Materese et al. 2015), and the resultant heterocyclic molecules are also of great biological significance with nitrogen-containing heterocycles such as pyrimidine and purine being the building blocks for nucleobases.

Pyrimidine has yet to be formally detected within the interstellar medium, however it possesses many similar physico-chemical characteristics to PAHs and so is expected to exist to some extent within icy grain mantles. The UV irradiation of frozen pyrimidine in the presence of $H_2O$, $CH_3OH$, $NH_3$, and $CH_4$ has been shown to yield the nucleobases uracil, cytosine, and thymine (Nuevo et al. 2009, 2012, 2014, Bera et al. 2016). Similarly, the irradiation of purine in ices containing $NH_3$ and $H_2O$ leads to the formation of the nucleobases guanine and adenine (Bera et al. 2015, Materese et al. 2017, 2018). Although the idea of an extended lifetime for these nucleobases within interstellar and near-Earth environments has been called into question (Peeters et al. 2003), if there exists some plausible mechanism by which they survive the stellar evolution phases of the astrochemical cycle depicted in Fig. 1, then it is possible that they may have been delivered to the early Earth via meteoritic or cometary impacts, thus providing the starting material for the so-called 'RNA World Hypothesis'.

The RNA World Hypothesis (Orgel 2004, Bernhardt 2012, Neveu et al. 2013, Higgs and Lehman 2015) hypothesizes that RNA or RNA-like molecules mediated the necessary information processing and metabolic transformations required for life to emerge from the early Earth's prebiotic environment. Systems based on RNA could act as a precursor to the significantly more complex system of RNA, DNA, and proteins on which current life is based, and it has been recognized that the ribonucleotide co-enzymes now used by many proteins may in fact be molecular fossils from an RNA-based metabolism (White 1976, Raffaelli 2011). The existence of naturally occurring ribozyme catalysts, such as self-splicing introns and the ribonuclease P catalyst (Kruger et al. 1982, Doudna and Cech 2002), along with the fact that ribosomal RNA actually catalyzes the formation of peptide bonds in the ribosome (Lang et al. 2008) has lent some weight to this hypothesis. However, much is still unknown as to the exact mechanisms by which life could arise in the RNA world, and so the hypothesis remains one among many seeking to explain the origins of biology from chemistry.

## 2    The Use of THz/F-IR Spectroscopy in Astrochemistry

THz/F-IR frequencies cover the region of the electromagnetic spectrum ranging between 0.1-10 THz (3-300 $cm^{-1}$). The main advantages of performing spectroscopy at such frequencies are the possibility of resolving ro-vibrational spectral transitions of gaseous molecules species, and the potential to probe long-range interactions between molecules in the solid phase in terms of low-energy intra- and intermolecular modes (Brown and Carrington 2003, Profeta and Scandolo 2011, Townes and Schawlow 2013). Thus, THz/F-IR spectroscopy makes for a unique tool in the detection of gases and ices in interstellar clouds and circumstellar disks against the widespread dust continuum via a number of distinct absorption and emission spectral signatures.

For example, the $H_2O$ molecule (whose energy transitions are well-defined) is an asymmetric rotor with an irregular set of energy levels characterized by the quantum numbers $J$ and $K$, where $J$ indicates the total rotational quantum number of the molecule and $K$ represents the projections of $J$ on the principal axes of inertia. Depending on the orientation of the nuclear spins of the constituent hydrogen atoms, the molecule is either ortho- (if spins of the hydrogen atoms are parallel) or para- (if the spins are anti-parallel). Each excitation or de-excitation of electrons from one energy level to another results in the propagation of photons of a certain energy (i.e., frequency). Fig. 4 shows the energy levels of ortho- and para-$H_2O$ in terms of their quantum numbers as detected by the Herschel Space Observatory. In this section, we provide a detailed description of the technical fundamentals of THz/F-IR spectroscopy, as well as an overview of its current applications to astrochemistry.

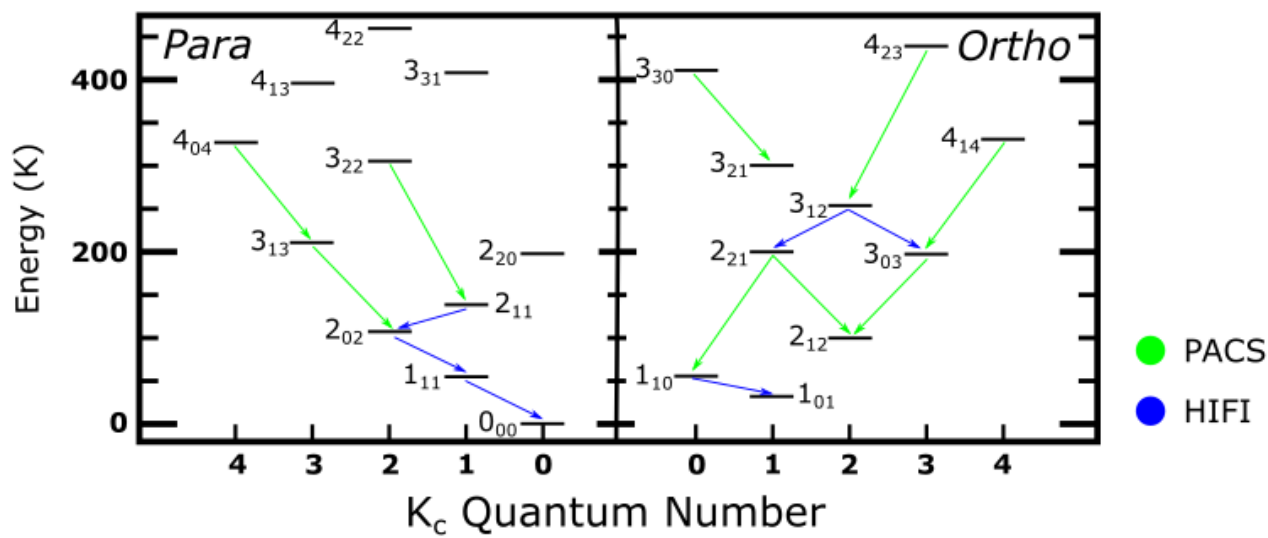

**Figure 4** Energy level diagram of ortho- and para-$H_2O$ showing the transitions observed by the PACS and HIFI instruments aboard the Herschel Space Observatory. Further information may be found in the work by van Dishoeck et al. (2013).

## 2.1 Spectrophotometer and Detector Technology

THz/F-IR laboratory spectroscopy measurements may be acquired using several different methods: Fourier-transform IR spectroscopy (Griffiths and de Haseth 1986), spectroscopy based on photo-mixing (Preu et al. 2011) or parametric conversion (Kawase et al. 1996, Kiessling et al. 2013), backward-wave oscillator spectroscopy (Medvedev et al. 2010), chirped-pulse Fourier-transform spectroscopy (Park and Field 2016), and time-domain spectroscopy (Auston et al. 1975, van Exter et al. 1989, Mittleman et al. 1998). These methods are all based on the detection of a continuous wave, broadband source, or pulsed THz/F-IR radiation illuminating gaseous or solid (ice) molecular samples of a species, and have varying frequency spectral ranges, sensitivities, and performances. For example, chirped-pulse spectrophotometers have a frequency bandwidth of up to 100 GHz (Park and Field 2016), while backward oscillation and time-domain spectrophotometers and have frequency ranges of up to 2.1 THz and 7.5 THz (Komandin et al. 2013, Allodi et al. 2014), respectively.

THz/F-IR time-domain spectroscopy can combine the generation of a terahertz pulse in a zinc telluride crystal via electro-optic rectification or from plasma filamentation in air using (in both cases) a femtosecond laser with a delay line and opto-electronic detection method in order to record the time domain trace (Allodi et al. 2013). This kind of spectroscopy has been widely used for gas-phase spectral line characterization in vacuum chambers (Kilcullen et al. 2015), in the study of refractive indices (Giuliano et al. 2019), and in the study of intermolecular transitions in astrophysical ice analogues, where the terahertz signal is reflected by or transmitted through the gas or ice. Although it is able to provide critical information regarding the frequency of the spectral transition and the absorption coefficient, it does have a few disadvantages including slow acquisition times and a fine-tuning requirement (Trofimov and Varentsova 2016). As such, these instruments are appropriate only for laboratory studies, and other detectors are usually found in observatories.

Incoherent or direct detector arrays are commonly used in THz/F-IR observatories combined with multiplexer readout systems. They require a higher sensitivity than the photon background noise (such as the microwave background) and high pixel counts. The main technologies used are bolometers (exploiting the neutron transmutation doping process or implanted silicon), superconducting transition-edge-sensed bolometers (TESBs), and kinetic inductance detectors (KIDs) (Farrah et al. 2019). Prior to the development of TESBs and KIDs, bolometers operating at temperatures of 300 K or lower were used in observatories. Incoming radiation results in a change in the detector temperature which is read out as a change in its resistivity. Standard bolometers, however, are limited in their total pixel number which has led to the development and preferential use of other technologies, including TESBs and KIDs.

A TESB consists of a superconducting film operating near its superconducting transitions temperature (Henderson et al. 2016, Thornton et al. 2016). The signal is seen as a current through a resistive film at a low temperature of around 100 mK (Suzuki et al. 2016). TESBs have high sensitivities which have allowed them to find use in many space-based instruments, as shown in Table 1. KIDs have a similar sensitivity to TESBs but operate below their superconducting transition temperature; below 300 mK (Griffin et al. 2016). The inductance of the detector material is increased due to broken Cooper pairs from incident photons on the superconducting film. In a resonant circuit, the shift in inductance causes a change in resonant frequency which may be detected (Day et al. 2005).

Coherent systems used in astronomy are based on the heterodyne detection principle: the THz/F-IR high-frequency (RF) signals from molecules in space are down-converted to a lower frequency band typically of a few GHz known as the intermediate frequency (IF). The IF is easily measurable by sensitive back-ends operating at low frequency (such as power meters, spectrum analyzers, digital spectrophotometers, etc.). The device used for detection and down-conversion is a mixer which requires the input of a local oscillator (LO) signal at the same frequency as the studied RF signal. LO generators need to have narrow linewidths, low noise, high stability, and a broad tunability with sufficient output power to couple to the mixer. In the THz/F-IR domain, multiple chains based on Schottky diodes and quantum cascade lasers (above 3 THz) are most commonly used (Valavanis et al. 2019).

There are three main types of mixers used in radio telescopes: superconductor-insulator-superconductor (SIS) diodes (Kojima et al. 2017), hot electron bolometers (HEBs) (Burke et al. 1999), and Schottky diodes (Maestrini et al. 2010). State of the art SIS diodes and HEBs offer the best system noise temperature performances at around 5-10 times the quantum noise limit. These mixers require only a few microwatts of LO power to operate and are very common for instruments with frequency ranges higher than 1 THz and between 1-6 THz for SIS diodes and HEBs, respectively. SIS diode mixers offer wider IF bandwidths and better sensitivities than do HEBs. However, SIS diodes require cooling to 4 K thus necessitating the use of a complex cryostat. On the other hand, Schottky mixers operate at temperatures above 70 K, a frequency range of up to 3 THz, and an IF bandwidth wider than 8 GHz. Their main disadvantages are their lower sensitivity which is between 30-50 times the quantum noise limit, and their milliwatt LO power requirement (Wilson et al. 2008).

The high-resolution spectroscopy possible when using a coherent detector in telescope facilities is achieved by using a digital technique with real-time fast Fourier-transform (FFT) so as to produce high-resolution spectral data. This idea was first proposed by Weinreb et al. (1961) and has been improved upon over the past 60 years in terms of acquisition speed, frequency bandwidth, power efficiency, cost, and size. Nowadays, three main digital systems are used in astronomy to sample the signal waveform at set time intervals: opto-acoustic spectrophotometers, digital autocorrelation spectrophotometers (Emrich 1997), and FFT spectrophotometers (Klein et al. 2012, Price 2016).

Opto-acoustic spectrophotometers are based on the diffraction of the signal on a Bragg crystal illuminated by a laser beam and detected by charge-coupled devices. This technique was used in the HIFI instrument aboard the Herschel Space Observatory (Siebertz et al. 2007). The main drawback of this spectrophotometer is its size, which defines its precision and spectral resolution. Digital autocorrelation spectrophotometers are based on the multiplication of a signal by a delayed version of the same signal using a series of delays. Spectrum measurement is achieved after the application of a Fourier transformation. Such spectrophotometers can have a bandwidth of a few GHz with a spectral resolution on the scale of one-hundredth of a kHz. Lastly, FFT spectrophotometers are based on field-programmable gate array (FPGA) chips. The FPGAs are combined with analogue-to-digital converters having a high data sampling rate of a few GHz samples per second. Their numerous spectral channels decompose the incoming RF signal into small sections giving an instantaneous FFT of a few kHz, as seen on the APEX telescope (Klein et al. 2006).

## 2.2    Telescope Facilities

Since the construction of the first radio telescope in 1937 (Kraus 1988), several telescopes operating in the THz/F-IR range have been successfully built and used by the astronomical community. A principal advantage of performing observations in the THz/F-IR range is the high sensitivity to low-abundance molecules within star-forming regions. Furthermore, observed emission and absorption features can be mapped for all areas within the stellar environment, such as snowlines where ice desorption processes begin to occur.

Progress in the sensitivity of detector technology over the past 50 years has permitted the investigation of interstellar molecular clouds at various wavelengths which has allowed us to improve our understanding of space chemistry. Powerful telescopes have been used with imaging and high spectral resolution instruments combined with large dish antennae. These telescopes have been instrumental in major scientific breakthroughs, such as the discovery of new molecules in the interstellar medium and elucidating the mechanisms for planetary system formation. A summary of the most important telescopes operating in the THz/F-IR range is given in Table 1.

In general, two varieties of observatories exist: atmosphere-based or space-borne telescopes. Atmosphere-based observatories are installed either on the ground or at high altitude or in airborne facilities. Atmospheric transmission has a direct impact on the sensitivity of these observatories at THz/F-IR frequencies since atmospheric molecules (especially $H_2O$) present strong absorption lines over these frequencies. As such, high atmospheric $H_2O$ content limits atmospheric transmission, thus making observation heavily dependent upon weather conditions and altitude: drier sites to install ground-based telescopes are found at higher altitudes (Fig. 5). Brief descriptions of the ground-based ALMA (Atacama Large Millimeter/Submillimeter Array) and the space-borne Herschel Space Observatory now follow.

ALMA is a ground-based international astronomy facility built over an area of 6596 m$^2$ at an altitude of 5000 m located at Llano de Chajnanto in the Atacama Desert, Chile. The 66 high-precision antennae making up the facility are either 12 m (54 of them) or 7 m (12 of them) in diameter. These sizes, combined with the area of the array, give a spatial resolution of 0.2-0.004 arcseconds (1 arcsecond =

1/3600 degrees). ALMA is composed of ten receivers using Schottky local oscillator sources and SIS diode mixers cooled to 4 K which give up to 16 GHz bandwidth with a spectral resolution of a few kHz. The covered frequency bands range from 31.3-950 GHz in ten windows (as shown in Fig. 5) with a maximum of 8 GHz IF. The lower frequency Bands 1 and 2 are not yet operational. The number of spectral channels available is 4096 per IF. For Bands 3-8, receivers operate in single sideband detection mode acquiring both H and V polarization while Bands 9 and 10 operate in double sideband mode.

**Table 1** Examples of the sensitivities achieved by selected observatories operating in the THz/F-IR range.

| Observatory | Aperture (m) | Instrument | Frequency Band (GHz) | Detector Technology |
|---|---|---|---|---|
| IRAM Telescope | 30 | NIKA2 | 150, 260 | KIDs |
| CSO (Caltech Submillimeter Observatory) | 10.4 | MUSIC | 142, 212, 272 | KIDs |
| | | SHARC II | 666, 857 | Bolometers |
| | | MAKO | 857 | KIDS |
| | | Heterodyne Receiver | 117-920 | SIS |
| | | Z-Spec | 190-230 | Bolometers |
| JCMT (James Clerk Maxwell Telescope) | 15 | SCUBA | 353, 666 | Bolometers |
| | | SCUBA2 | 353, 666 | TES |
| | | HARP | 325-375 | SIS |
| | | Nāmakanui | 86, 215-371 | SIS |
| ODIN | 1.1 | SMR | 118-580 | Schottky |
| SMA (Submillimeter Array) | 6×6 | Band 1-4 | 180-700 | SIS |
| APEX (Acatama Pathfinder Experiment) | 12 | ArTeMis | 666, 856 | Bolometers |
| | | SEPIA | 157-738 | SIS |
| | | nFLASH | 196-507 | SIS |
| | | CHAMP | 620-850 | SIS |
| | | PI230 | 157-211 | SIS |
| | | LABOCA | 345 | Bolometers |
| NEOMA interferometer | 10×15 | Band 1-4 | 71-371 | SIS |
| Herschel Space Observatory | 3.5 | SPIRE | 500-1500 | Bolometers |
| | | PACS bol. | 1427-5000 | Bolometers |
| | | PACS phot. | 1363-6000 | Photoconductors |
| | | HIFI | 83-1000 | Bolometers |
| SOFIA (Stratospheric Observatory for Infrared Astronomy) | 2.5 | HAWC+ | 1250-6000 | TES |
| | | FIFI-LS | 1500-5880 | Photoconductors |
| | | GREAT | 490-4745 | HEB |
| LMT (Large Millimeter Telescope) | 50 | ToITEC | 150, 220, 280 | KIDs |
| | | SEQUOIA | 85-116 | SIS |
| | | AzTEC | 272 | Bolometers |
| | | RSR | 75-110 | SIS |
| ALMA (Atacama Large Millimeter/Sub-millimeter Array) | 54×12 12×7 | Band 3-10 | 84-950 | SIS |
| Millimetron | 10 | Band 1-2 | 33-116 | HEMT |
| | | Band 3-5 | 211-650 | SIS |
| OST (Origins Space Telescope) | 5.9 to 9.1 | Imaging | 1000-3000 | - |
| | | Spectroscopy | 1000-3000 | - |

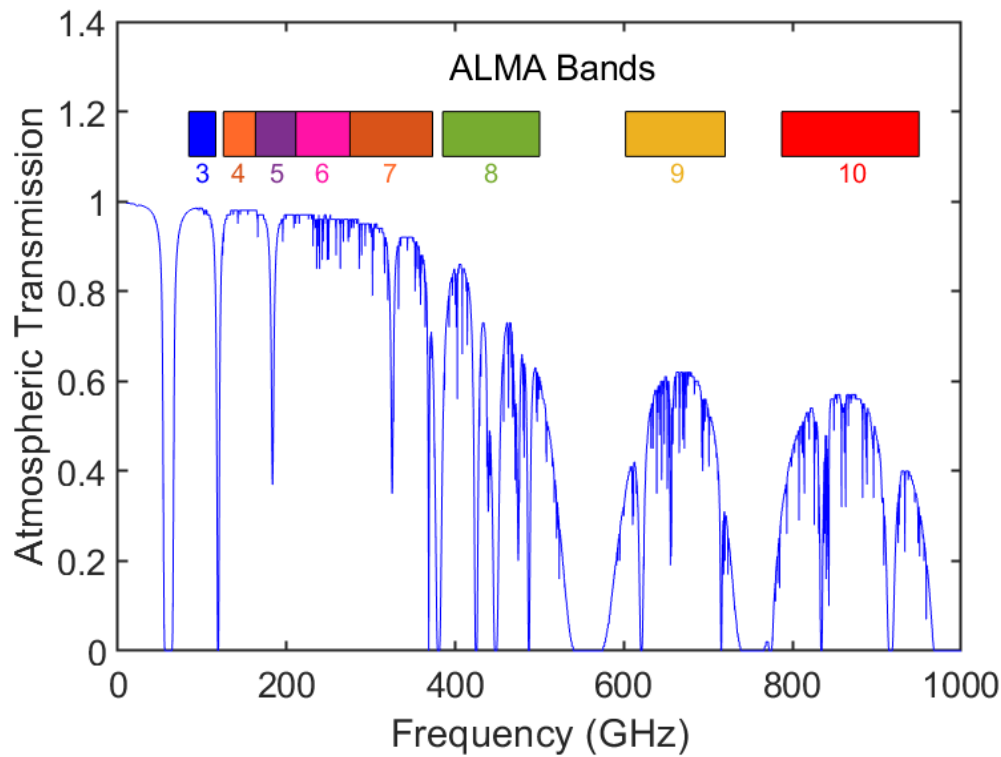

**Figure 5** Zenith atmospheric transmission of $H_2O$ vapor of the ALMA telescope with precipitable $H_2O$ vapors of 0.5 mm simulated from the *Atmospheric Transmission at Microwaves* model described by Pardo et al. (2001). Further information may be found in the work by Maiolino (2008).

Space-based telescopes are not limited by atmospheric attenuations in the way that atmospheric observatories are, but do have limitations connected with the size of their antenna aperture. The Herschel Space Observatory was commissioned by the European Space Agency and launched in 2009. The telescope was composed of a 3.5 m diameter dish which was passively cooled by liquid helium at 4 K. By 2013, the liquid helium reserves necessary for detector cooling had been gradually depleted, defining the end of its operations in April 2013.

Three scientific instruments were on board the Herschel Space Observatory: a high spectral resolution heterodyne spectrophotometer called HIFI (Heterodyne Instrument for the Far-Infrared), an imaging photometer called PACS (Photodetector Camera Array), and a medium resolution grating spectrophotometer called SPIRE (Spectral and Photometric Imaging Receiver). THz/F-IR frequencies were covered by the HIFI instrument from 408-1908 GHz separated in 12 windows using HEB and SIS diode mixer technology. PACS was composed of a camera integral field spectrophotometer combined with an imaging photometer operating between 1.427-6 THz and bolometers at frequencies centered around 1.873 THz, 2.997 THz, and 4.282 THz (Rosenthal et al. 2002). Finally, the SPIRE instrument had three main photometers at 600 GHz, 856 GHz, and 119 GHz, as well as a two-band imaging Fourier-transform spectrophotometer with bolometers operating between 447-1550 GHz (Griffin et al. 2010).

Amongst the key scientific objectives of the Herschel Space Observatory was an investigation of the formation of stellar and planetary systems, and to increase our understanding of the physics and chemistry of the interstellar medium, our Solar System, and extra-galactic galaxies. The HIFI instrument detected more than 100,000 spectral features in a single spectrum (Bergin et al. 2010). Through the Water in Star Forming Regions with Herschel (WISH) program, many molecules such as $H_2O$, CO, and $O_2$ were detected in various astrophysical environments (Goldsmith et al. 2011). Molecules and radicals including $H_2O$ and $OH^-$ were detected further out and deeper in protoplanetary disks than before from surveys of THz/F-IR spectral lines using PACS (Fedele et al. 2013). Cold water reservoirs were discovered in two extra-solar planet-forming disks (TW Hydrae) through the detection of $H_2O$ gas produced by UV photo-desorption of the ice made by HIFI (Hogerheijde et al. 2011). $H_2O$ vapor was also detected in a pre-stellar core at the early stage of stellar formation using HIFI observations combined with models of the UV photo-desorption processes (Caselli et al. 2012).

Atmosphere-based and space-borne observatories operating in the THz/F-IR range, including those listed in Table 1, have aided astronomers in elucidating the chemistry of interstellar space by allowing them to obtain spectroscopic information of molecular species (colloquially termed 'molecular fingerprints'). Indeed, the improvement of technology throughout these past decades has allowed the definitive identification of many new molecular signatures. Fig. 6 shows the comparison of spectra acquired by ALMA Band 10 and the HIFI instrument aboard the Herschel Space Observatory while observing the NGC 6334I region (McGuire et al. 2018b). As can be seen, the number of spectral lines attributed to CO and $CH_3OH$ which are resolved by ALMA Band 10 is significantly higher than that detected by HIFI (Zernickel et al. 2012).

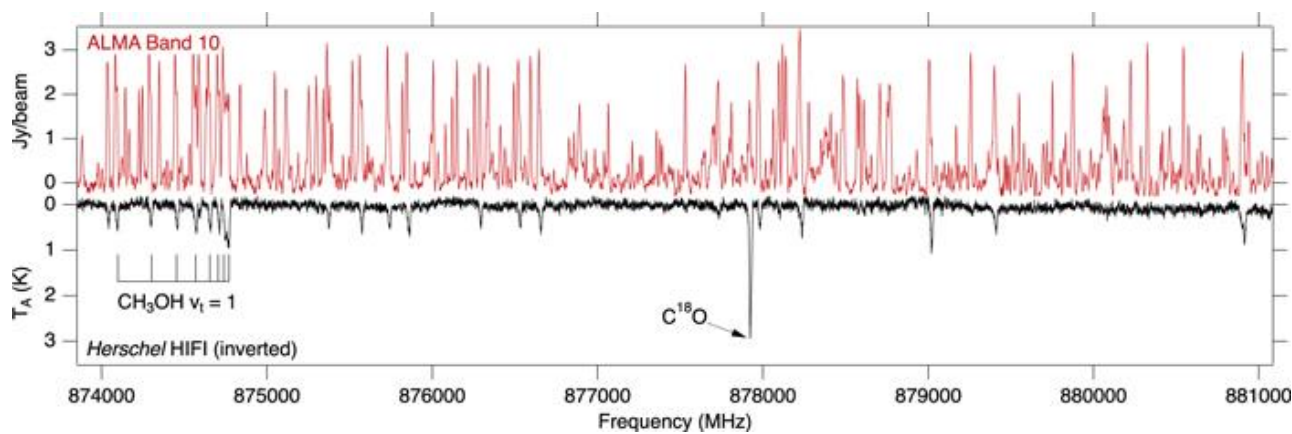

**Figure 6** Comparison of the spectra acquired by ALMA Band 10 and HIFI while observing the NGC 6334I region. Figure taken from McGuire et al. 2018b and reproduced by permission of the AAS.

Computer modelling and laboratory measurements of molecular spectra are often needed to assign observed spectral lines to particular molecular species. This is especially true in the case of complex organic molecules such as isopropyl cyanide $((CH_3)_2CHCN)$, which was the first branched carbon-chain molecule to be detected and which was found in Sgr B2 using the ALMA Band 3 (Belloche et al. 2014). Later, $CH_3OCH_2OH$ and $CH_3Cl$ were observed for the first time by ALMA Bands 6 and 7 in NGC 6334I, and by ALMA Band 7 in IRAS 16293-2422, respectively (Fayolle et al. 2017, McGuire et al. 2017). However, many molecules and radical species may not be observed in the gaseous phase but may play a central role in the formation of observed gaseous species by their presence and reactions in the solid phase in interstellar icy grain mantles. The use of laboratory THz/F-IR measurements in astrochemical research as it pertains to solid-phase chemistry will be elaborated on in the next section.

### 2.3 Laboratory Astrochemistry Using THz/F-IR Spectroscopy

Over 200 molecules have been detected and spectroscopically resolved in interstellar regions within our galaxy, and more than 60 have been observed in extra-galactic molecular clouds (Müller et al.

2005), ranging in complexity from simple diatomics to the more structurally complex fullerenes and PAHs. An overarching aim of laboratory astrochemistry is to be able to understand the physico-chemical processes which occur in these astrophysical environments so as to better comprehend the chemical reactions which lead to the formation of such molecules. Such work is motivated by the knowledge that such reactions can result in the production of various molecules and minerals. For instance, interstellar biomolecules which may form in dense, quiescent molecular clouds may be subsequently incorporated into the stellar and planetary systems which evolve from these clouds and thus play a defining role in the emergence of life within these systems (Watanabe et al. 2002, Jones et al. 2011, Fedoseev et al. 2017, Sandford et al. 2020, Chuang et al. 2020, Ioppolo et al. 2021). The synthesis of smaller, volatile molecules in astrophysical environments is equally important, as such molecules may contribute to the development of planetary and lunar atmospheres or transient exospheres (Milillo et al. 2016, Teolis and Waite 2016). Additionally, condensation processes and grain alterations contribute to the formation of minerals which are significant to the geological and geochemical evolution of planets and moons, including graphite, corundum, moissanite, and forsterite (Hazen et al. 2008, Hazen and Ferry 2010).

The unambiguous identification of such molecules by radio telescopes relies on the availability and completeness of spectroscopic gas- and solid-phase databases at the corresponding frequencies for a great variety of molecules. Such information is obtained by performing classical laboratory absorption measurements (Linnartz et al. 2010). Presently, the two most commonly referenced catalogues are the Cologne Database for Molecular Spectroscopy (Müller et al. 2001, Endres et al. 2016) and the Jet Propulsion Laboratory (Pickett et al. 1998) catalogues, which give the frequency and amplitude of several molecular transitions. Two accessible databases listing vibrational transitions at IR wavelengths are also available: the HITRAN database (Rothman et al. 2009) and the EXOMOL line list.

Broadly speaking, laboratory astrochemistry studies are largely concerned with investigating reactions which occur in either the gas phase or the condensed (ice) phase. Experiments are typically performed using high or ultra-high vacuum chambers. For instance, a novel method of performing gas-phase spectroscopy is to make use of a heterodyne radiometer so as to observe the absorption features of various gaseous species within the chamber, thus functioning in a similar fashion to several telescope facilities. Two laboratory broadband emission THz/F-IR radiometers based on heterodyne detection methods have been developed: the first is based on Schottky-barrier technology combined with a FFT digital spectrophotometer operating between 80-110 GHz (Wehres et al. 2017), while the second is based on cryo-cooled SIS diode mixer technology operating between 270-290 GHz (Wehres et al. 2018). By using such instruments, the spectroscopic features of pyridine and $CH_3CN$ have been successfully measured and matched with analytical simulations. Similarly, a Schottky-based receiver has been used in combination with a FFT digital spectrophotometer and vacuum chamber to observe the emission line of $N_2O$ at 355.6 GHz (Parkes et al. 2018).

Laboratory experiments concerned with the solid phase also require the use of high or ultra-high vacuum chambers, wherein interstellar ice analogues may be synthesized on cold substrates via the direct or background deposition of dosed gases or vapors, or the effusive evaporation of refractories. The deposited ices are then processed in such a way as to simulate astrophysical conditions: atom (or free radical) additions represent non-energetic pathways towards molecule formation (Linnartz et al. 2015, Chuang et al. 2016, Potapov et al. 2017, Ioppolo et al. 2021); bombardment with energetic charged particles (i.e., electrons and ions) simulates ice interactions with galactic cosmic rays, the solar wind, or planetary magnetospheric plasmas (Baragiola et al. 1999, Dalton et al. 2013, Ding et al. 2013, Boamah et al. 2014, Boduch et al. 2015, Fulvio et al. 2019, Ioppolo et al. 2020); irradiation with photons (both ionizing and non-ionizing) simulates extra-terrestrial photochemistry (Bernstein et al. 2012, Lo et al. 2014, Öberg 2016, Mullikin et al. 2018); energetic shocks simulate collisions between interstellar or Solar System bodies (Goldman et al. 2010, Martins et al. 2013), and thermal processing simulates the chemical and structural changes induced within interstellar icy grain mantles proximal to nascent stars, or in Solar System bodies as they approach perihelion (Ehrenfreund et al. 1999, Kaňuchová et al. 2017).

The changes in the physico-chemical properties of the ice analogues induced by such processing have been traditionally monitored using mid-IR or UV-VIS spectroscopy (Boersma et al. 2014). This has been advantageous in that such laboratory work is complementary to observational work using space-borne telescopes working within these ranges of the electromagnetic spectrum, such as the Spitzer Space Telescope and the Hubble Space Telescope. Although not yet widely used within laboratory settings, the utility and applicability of THz/F-IR spectroscopy is being increasingly recognized due to its ability to detect lower frequency molecular vibrations, as well as intermolecular interactions and translational vibrations within solid lattices (Bertie 1968, Bertie and Jacobs 1977, Kulesa 2011, McIntosh et al. 2012).

THz/F-IR spectra differ depending upon the structural morphology of the ice (e.g., amorphous or crystalline solid phases), as demonstrated by Moore and Hudson (1992, 1994, 1995). For instance, recent studies investigating $CO_2$ in $H_2O$ and $CH_3OH$ ice matrices at various temperatures have revealed that the measured THz/F-IR spectra are especially sensitive to the degree of segregation within the ice structure. Results have shown that $CO_2$ roaming during warming of the ice results in a localized accumulation, or segregation. When mixed with crystalline $H_2O$ ice, this $CO_2$ segregation results in a disruption of the hydrogen bonding network between adjacent $H_2O$ molecule layers, causing the $H_2O$ spectral resonance features to be shifted, distorted, or attenuated (Allodi et al. 2014). On the other hand, $CO_2$ absorption bands were observed to become narrower and more defined when segregation from $CH_3OH$ ice occurred as a result of warming (McGuire et al. 2016).

Such work clearly demonstrates the sensitivity of THz/F-IR spectroscopy not only to the nature of the chemical species present within the ice, but also to the physical structure of the ice. Given that the nature of the observed spectral absorption features is dependent upon the temperature, degree of crystallinity, segregation, and thermal history of the ice, THz/F-IR spectroscopy appears to be highly suitable for assessing not only the composition and temperature characteristics of astrophysical ices, but also their morphology (Ioppolo et al. 2014). To date, relatively few of the observed molecules have been investigated using THz/F-IR spectroscopy. Amorphous solid $H_2O$, for instance, presents a spectral line at 1 THz. $CH_3OH$, $CH_3CHO$, $(CH_3)_2CO$ have been found to have a broad feature at around 4 THz, while in HCOOH and $CH_3COOH$ this broad absorption feature is centered at about 7 THz (Ioppolo et al. 2014).

Laboratory THz/F-IR spectroscopy has also been used in the analysis of more complex molecular structures such as PAHs. Although these molecules are believed to be ubiquitous in interstellar and circumstellar media (Tielens 2008, Kwok and Zhang 2011), few single structures have been positively identified. Systematic analysis of a series of PAHs, as well as their hydrogenated and alkylated derivatives, has revealed that the THz/F-IR spectra of these molecules display so-called 'Jumping Jack' modes which correspond to the in-plane vibrations around the central molecular core (Cataldo et al. 2013). As such, these spectra provide crucial information related to the number of fused aromatic rings present within the molecule, and are thus valuable in identifying the individual molecular carriers of PAH diffuse interstellar absorption bands. Other complex molecules, including those which may be of relevance to prebiotic chemistry, have also been studied using THz/F-IR spectroscopy (Widicus-Weaver et al. 2005, Carroll et al. 2010).

Recently, a new laboratory-based method for observing desorption mechanisms occurring within the interstellar medium and star-forming regions has been developed. Studying such desorption has been traditionally performed using mid-IR spectrophotometers and quadrupole mass spectrometers, however this has the drawback of not being able to detect ions and radicals and is difficult to relate to observations made using radio telescopes (Fraser et al. 2002, Collings et al. 2003, Fuchs et al. 2009, Muñoz-Caro et al. 2010). In principle, it is possible to desorb (either via thermal or non-thermal mechanisms) ices formed of pure or layered molecules and observe the resultant gas phase in the THz/F-IR frequency range. One way to do this is to passively observe the emission of those molecules during desorption in an astrochemical chamber equipped with a heterodyne radiometer (Auriacombe et al. 2015, 2016); another is by performing microwave absorption spectroscopy. The region in which desorbing molecules are observed may be directly above the ice, and the relevant spectral transitions may be detected with a HEB (see Fig. 2 in Yocum et al. 2019). Alternatively, a chirped-pulse Fourier-transform spectrophotometer can be used to observe the desorption occurring inside a waveguide (Theulé et al. 2020). These emerging techniques are capable of bringing new insights and possibilities to our understanding of desorption processes in star-forming regions.

Although the use of THz/F-IR spectroscopy is growing within laboratory-based astrochemical research (driven by developments in the relevant technologies and an increase in the commercial availability of the necessary spectrophotometers), much of the current work takes place at large-scale facilities. A complete and thorough description of the research conducted using terahertz radiation at such facilities would go beyond the scope of this review, and we instead direct the interested reader to other works (e.g., Neil 2014). However, it is worthwhile to take some time to briefly describe the Free Electron Lasers for Infrared Experiments (FELIX) facility located in Nijmegen, the Netherlands, where a high-vacuum chamber has been installed as an end-station thus permitting the characterization of astrophysical ice analogues which have been processed as a result of THz/F-IR irradiation.

The FELIX facility is home to four beamlines which provide short-pulsed coherent light covering the microwave to mid-IR regions of the electromagnetic spectrum. The FELIX-1 and FELIX-2 free electron laser beamlines respectively operate over the ~65-330 $cm^{-1}$ and ~225-3300 $cm^{-1}$ ranges and are capable of providing bursts of micro-pulses with selective and tunable wavelengths whose duration may be readily controlled. These beamlines are therefore highly suited to providing THz/F-IR (as well as mid-IR) radiation for use in laboratory studies of astrophysical ice analogues, which are prepared in the Laboratory Ice Surface Astrophysics (LISA) end-station.

LISA is a high-vacuum chamber with nominal base pressure in the region of a few $10^{-9}$ mbar containing an elongated rectangular gold-coated deposition substrate which may be cooled down to 15 K using a closed-cycle helium cryostat head with a cooled compressor. The gold-coated substrate is manipulated by means of a *xyz*-linear translator allowing for the irradiation of different spots of the ice during a single experiment, as well as a rotary stage mostly employed for alignment purposes. Pure gases and gas mixtures are pre-prepared in a dosing line prior to being injected into the main chamber via an all-metal leak valve, wherein background deposition of the gases occurs to form the astrophysical ice analogues. Such background deposition of gases onto the deposition substrate allows for a more uniform ice deposition, and the resultant ice analogue may be used in systematic studies. Ices processed by mid-IR or THz/F-IR irradiation delivered by the FELIX beamlines may be monitored in two ways: mid-IR absorption spectroscopy in reflection mode may be used to monitor any physico-chemical effects induced as a result of the ice stimulation, and quadrupole mass spectrometry may be used to monitor species which desorb thermally from the bulk parent ice.

The FELIX-LISA facility has found regular use in effectively characterizing the physico-chemical changes induced in astrophysical ice analogues as a result of stimulation by incident irradiation. For instance, Noble et al. (2020) investigated the effect of resonant irradiation of amorphous solid $H_2O$ using mid-IR radiation supplied by the FELIX-2 beamline. Their results indicated that irradiation with wavelengths corresponding to the mid-IR stretching, bending, and libration modes of amorphous $H_2O$ resulted in a wide-ranging structural rearrangement of the ice into a crystalline-like form due to vibrational relaxation of the intermolecular hydrogen bonding network. Results such as these, combined with the experimental capabilities of large-scale facilities such as FELIX, make future dedicated studies looking into the irradiation of lower frequency THz/F-IR modes of astrophysical ice analogues especially attractive.

## 3    Conclusions and Recommendations for Future Work

As has been demonstrated, the application of THz/F-IR spectroscopy to observational and laboratory astrochemistry has the potential to unravel much information related to the structure and reactivity of molecules in space. To date, however, THz/F-IR experiments at large-scale facilities such as FELIX tasked with investigating the chemistry of interstellar and Solar System ice analogues have not been explored to their fullest potential due to the fact that dedicated pump-probe experiments required to study such chemistry are not available at the time of writing. The present LISA end-station configuration comprises an FTIR spectrophotometer with an extended spectral range to the far-IR. Such a system may be used to time-resolve transient events within the ice structure (such as diffusion) at millisecond timescales.

To fully benefit from the unique capabilities of a free electron laser (i.e., a wide-range tunability, high peak power, and a controllable repetition rate), terahertz time-domain spectroscopy with single-shot detection techniques should be employed to extend time resolutions to a few tens of picoseconds. With such a configuration, free electron laser THz/F-IR radiation could be used to 'pump' (i.e., inject energy into) inter- and intramolecular vibrations of solid species that, having been excited by the sudden excess of energy, will subsequently rearrange selectively or diffuse within the ice and possibly react with other species. Diffusion and reaction of molecules could be triggered and controlled by operating a free electron laser and could then be monitored by means of a terahertz time-domain system in the solid phase, while desorbed species could be monitored by implementing a broadband emission THz/F-IR radiometer as described previously. The combination of a free electron laser and new advanced techniques in laboratory astrochemistry thus have the potential to reveal unprecedented details on fundamental phenomena which could play important roles in the formation of complex prebiotic molecules in space, and would thus expand our understanding of the relevant physics and chemistry.

Experimental work such as that proposed to be performed at the FELIX-LISA facility would also contribute much to our interpretation of observational data. Interestingly, the far-IR is one of the spectral ranges where astronomical data with sub-arcsecond resolution is not currently available; a fact which is somewhat at odds with the potential usefulness of such spectroscopic techniques (Linz et al. 2021). A number of space-based observatories, such as the FIRSS (Far-Infrared Spectroscopic Surveyor) and the THEZA (Terahertz Exploration and Zooming-In for Astrophysics) telescopes have been proposed to fill in this gap (Gurvits et al. 2021, Rigopoulou et al. 2021), and the acquisition of laboratory reference data will doubtlessly aid in further characterizing the molecular composition of the cosmos.

In summary, we have provided a brief overview of the chemistry occurring in both diffuse and dense (dark) interstellar clouds with a focus on ion-molecule reactions in the gas and solid phases which are relevant to the formation of complex prebiotic molecules, as well as how THz/F-IR techniques have aided in the elucidation of these reactions and how such reactions may relate to the RNA World Hypothesis on the abiogenic origins of life. We have also provided a detailed review on current THz/F-IR spectrophotometer and detector technology in both laboratory and observatory settings, as well as their present and potential future applications to astrochemistry. We conclude by emphasizing that further progress in THz/F-IR astrochemistry has the potential to provide great insight to several fundamental astrochemical processes and should therefore be pursued.

## 4    Conflict of Interest

The authors declare that the research was conducted in the absence of any commercial or financial relationships that could be construed as a potential conflict of interest.

## 5    Author Contributions

DVM, PAH, ATM, OA, and SI wrote the manuscript and all authors were responsible for corrections and improvements.

## 6    Funding

Our research has benefitted from support provided by the Europlanet 2024 RI, which has received funding from the European Union's Horizon 2020 research innovation program under grant agreement No. 871149.

## 7    Acknowledgments

DVM is the grateful recipient of a University of Kent Vice-Chancellor's Research Scholarship. ATM thanks Queen Mary University of London for doctoral funding. SI acknowledges the Royal Society for financial support.

## 8    ORCID ID Numbers

Perry A. Hailey:                0000-0002-8121-9674

Duncan V. Mifsud:                0000-0002-0379-354X

Alejandra Traspas Muiña:                0000-0002-4304-2628

Olivier Auriacombe:                0000-0002-5810-8650

Sergio Ioppolo:                0000-0002-2271-1781

Nigel J. Mason:                0000-0002-4468-8324